\documentclass[conference]{IEEEtran}
\usepackage{amsfonts}
\usepackage{graphicx}
\usepackage{amsmath}
\usepackage{subfigure}
\usepackage{multirow}
\usepackage{makecell}
\usepackage{algorithm}
\usepackage{algorithmic}
\usepackage{amsthm}
\usepackage{bm}
\usepackage[OT1]{fontenc}

\usepackage[square, comma, sort&compress, numbers]{natbib}

\usepackage{tikz,pgfplots} 
\usepackage{pgfplotstable} 
\usepackage{textcomp}
\newcounter{groupcount}
\pgfplotsset{
	draw group line/.style n args={5}{
		after end axis/.append code={
			\setcounter{groupcount}{0}
			\pgfplotstableforeachcolumnelement{#1}\of\datatable\as\cell{%
				\def\temp{#2}
				\ifx\temp\cell
				\ifnum\thegroupcount=0
				\stepcounter{groupcount}
				\pgfplotstablegetelem{\pgfplotstablerow}{[index]0}\of\datatable
				\coordinate [yshift=#4] (startgroup) at (axis cs:\pgfplotsretval,0);
				\else
				\pgfplotstablegetelem{\pgfplotstablerow}{[index]0}\of\datatable
				\coordinate [yshift=#4] (endgroup) at (axis cs:\pgfplotsretval,0);
				\fi
				\else
				\ifnum\thegroupcount=1
				\setcounter{groupcount}{0}
				\draw [
				shorten >=-#5,
				shorten <=-#5
				] (startgroup) -- node [anchor=north] {#3} (endgroup);
				\fi
				\fi
			}
			\ifnum\thegroupcount=1
			\setcounter{groupcount}{0}
			\draw [
			shorten >=-#5,
			shorten <=-#5
			] (startgroup) -- node [anchor=north] {#3} (endgroup);
			\fi
		}
	}
}

\pgfplotsset{grid style={step=1.0, dotted,black}} 
\pgfplotsset{compat=newest} 

\DeclareMathOperator*{\argmin}{arg\,min}

\pgfplotstableread{
	1   1792	1792	1792	1
	2   616		422		1104  	1
	3   4096	4096	4096	2
	4   1444	956		2632  	2
}\datatable

\begin{document}
	
	\title{Polar Decoding on Sparse Graphs with Deep Learning}
	
	\author{\IEEEauthorblockN{Weihong Xu$^{1,2}$, Xiaohu You$^{2}$, Chuan Zhang$^{1,2}$ and Yair Be'ery$^3$}
		\IEEEauthorblockA{$^{1}$Lab of Efficient Architecture for Digital Communication and Signal Processing (LEADS)\\
			$^{2}$National Mobile Communications Research Laboratory, Southeast University, Nanjing, China\\
			$^{3}$School of Electrical Engineering, Tel-Aviv University, Israel
			\\
			Email: $^2$\{wh.xu, xhyu, chzhang\}@seu.edu.cn, $^3$ybeery@eng.tau.ac.il}
	}
	
	\maketitle
	
	\begin{abstract}
		In this paper, we present a sparse neural network decoder (SNND) of polar codes based on belief propagation (BP) and deep learning. At first, the conventional factor graph of polar BP decoding is converted to the bipartite Tanner graph similar to low-density parity-check (LDPC) codes. Then the Tanner graph is unfolded and translated into the graphical representation of deep neural network (DNN). The complex sum-product algorithm (SPA) is modified to min-sum (MS) approximation with low complexity. We dramatically reduce the number of weight by using single weight to parameterize the networks. Optimized by the training techniques of deep learning, proposed SNND achieves comparative decoding performance of SPA and obtains about $0.5$ dB gain over MS decoding on ($128,64$) and ($256,128$) codes. Moreover, $60 \%$ complexity reduction is achieved and the decoding latency is significantly lower than the conventional polar BP.
	\end{abstract}


	\section{Introduction}\label{sec:introduction}

	Deep neural network (DNN) and deep learning techniques show promising performance in vast variety of tasks. In quantum information, DNN is utilized to decode stabilizer code \cite{krastanov2017deep} through encoding the probability distribution of errors. The authors in \cite{ye2017initial} discuss the possibility of applying DNN to channel equalization and decoding. Recurrent neural network (RNN) is adopted to detect data sequences \cite{farsad2018neural} in communication systems.
		
	On the other side, polar codes \cite{arikan2009channel} are regarded as a prominent breakthrough in channel coding because of their capacity-achieving property. Now polar codes have been selected as the error-correcting codes of the enhanced mobile broadband (eMBB) control channels for the 5th generation (5G) wireless communication systems. With the advanced deep learning libraries and high performance hardware, many efforts have been made to develop a neural network decoder (NND) that can adaptively decode polar codes under different channel conditions.	The authors in \cite{gruber2017deep} exploit naive dense neural network to decode very short polar codes. It shows that NND trained by all possible codewords leads to near maximum a posteriori (MAP) performance. But the complexity is prohibitive due to the exponential nature of binary codewords. To alleviate the enormous complexity of long polar codes, \cite{cammerer2017scaling} partitions the polar encoding graph into small blocks and train them individually. Although the degradation of partitioning is negligible, the overall decoding complexity is still high. To overcome these issues, in \cite{xu2017improved}, trainable weights are assigned to the edges of belief propagation (BP) factor graph and then the iterative BP decoding is converted into DNN. The method requires much lower complexity and less parameters compared to \cite{gruber2017deep, cammerer2017scaling}, which is feasible for long polar codes. However, the decoding latency is long since the depth of NND is determined by iteration number and code length.
	
	In this work, we propose a sparse neural network decoder (SNND) for polar codes with high parallelism, low latency and low complexity. Inspired by \cite{nachmani2016learning}, our SNND is constructed from the bipartite Tanner graph of polar codes in \cite{cammerer2017sparse}. The sum-product algorithm (SPA) is replaced by min-sum (MS) approximation to reduce complexity. After the network is trained by deep learning techniques, SNND achieves the equal bit error rate (BER) performance with SPA decoding. Moreover, the decoding latency is about ${1}/{\log_{2}N}$ of the conventional polar BP \cite{arikan2008performance} due to the fully parallel structure.
		
	The remainder of this paper is organized as below. Polar codes and BP decoding are briefly introduced in Section \ref{sec:preliminaries}. Section \ref{sec:proposed_snnd} describes how to construct the sparse trellis of SNND. Then the corresponding decoding process and model training methodology are given in detail. The experiment results in Section \ref{sec:experiment} demonstrate the improvements of proposed SNND over various code lengths. The latency and complexity analysis is also given. Section \ref{sec:conclusion} concludes this paper.

	\section{Preliminaries}\label{sec:preliminaries}
	
	\subsection{Polar Codes}
	Polar codes have proven to be capable of achieving the capacity of symmetric channel \cite{arikan2009channel}. The encoder of an ($N,K$) polar code assigns $K$ information bits and the other ($N-K$) bits to the reliable and unreliable positions of the $N$-bit codeword $\mathbf{u}^{N}$, respectively. Those bits in unreliable positions are referred as frozen bits and usually fixed to zeros. Then, the $N$-bit transmitted codeword $\mathbf{x}^{N}$ can be obtained according to $\mathbf{x}^{N} = \mathbf{u}^{N} \mathbf{G}_{N}$, where $\mathbf{G}_{N}$ is the generator matrix and satisfies $\mathbf{G}_{N} = \mathbf{F}^{\otimes n}$. Note that $\mathbf{F}^{\otimes n}$ is the $n$-th \textit{Kronecker} power of $\mathbf{F}= \resizebox{.1\hsize}{!}{$
		\begin{bmatrix}
			1 & 0 \\
			1 & 1
		\end{bmatrix}$}$ and $n = \log_{2}N$.
	
	\subsection{Belief Propagation Decoding}
\begin{figure}[ht]
		\centering
		\includegraphics[width=0.5\linewidth]{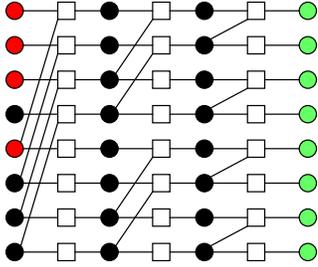}
		\caption{Factor graph of (8,4) polar code with $\mathcal{A} = \{4,6,7,8\}$ \cite{cammerer2017sparse}.}
		\label{fig:polar_BP_fg}
	\end{figure}

	BP is one of the commonly used message passing algorithms for polar decoding. The BP algorithm decodes polar codes through iteratively processing the log-likelihood ratios (LLRs) over the factor graph of any ($N,K$) polar code. Unlike the fully parallel Tanner graph of LDPC codes, the factor graph of polar decoding is based BP decoder for Reed-Muller (RM) codes. In this case, the factor graph consists of $n = \log_{2}N$ stages and $(n+1)N$ nodes in total. Fig. \ref{fig:polar_BP_fg} illustrates the factor graph of ($8,4$) polar code.
	
	\subsection{LDPC-like Polar Decoder}
	
	The polar BP decoder is generally constructed based the generator matrix $\mathbf{G}_{N}$, which has a similar trellis structure with its encoding factor graph. However, this causes inefficiencies for polar decoding since the number of stages is determined by the code length. Moreover, the multiple-stage architecture of polar decoder results in longer latency compared with the fully parallel scheduling of LDPC-like BP decoding.
	
	\begin{figure}[ht]
		\centering
		\subfigure[Dense]
		{
			\includegraphics[width=0.15\linewidth]{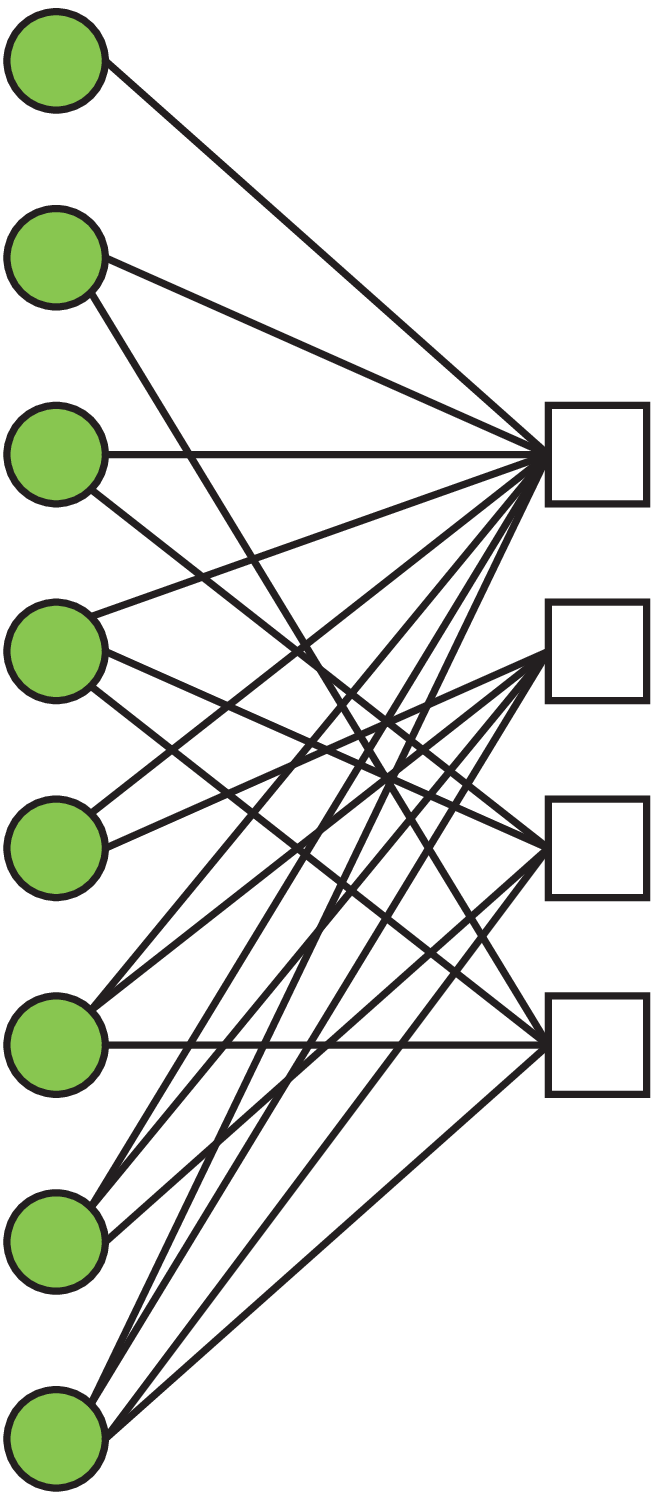}
			\label{fig:dense_tanner}
		}
		\hfil
		\subfigure[Sparse]
		{
			\includegraphics[width=0.15\linewidth]{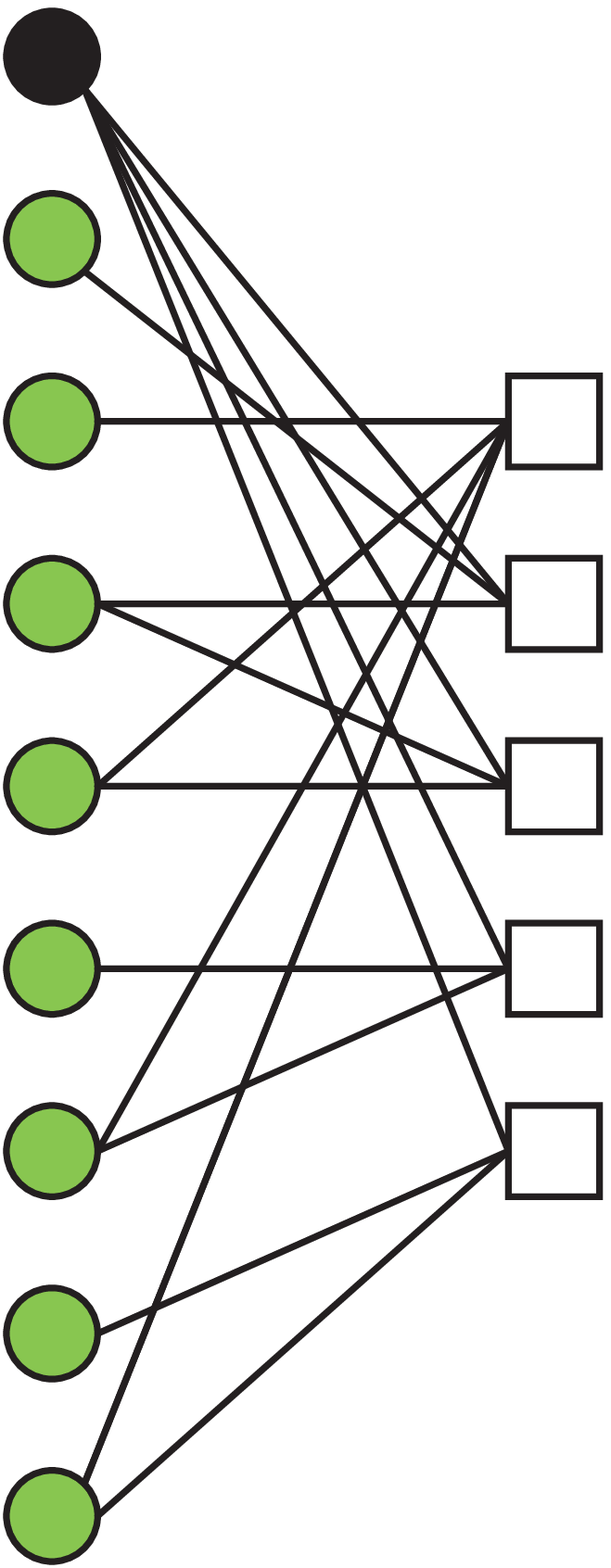}
			\label{fig:sparse_tanner}
		}
		\caption{LDPC-like Tanner graphs \cite{cammerer2017sparse} for (8,4) polar code with $\mathcal{A} = \{4,6,7,8\}$.}
		\label{fig:graph}
	\end{figure}

	To overcome the aforementioned problems, the parity-check matrix $\mathbf{H}$ of polar codes can be constructed from the corresponding generator matrix $\mathbf{G}_{N}$ in \cite{goela2010lp}. The conventional polar BP factor graph is then converted to the LDPC-like bipartite graph (see Fig. \ref{fig:dense_tanner}) consisting of variable nodes (VNs) and check nodes (CNs). But the dense graph representation involved with many circles has demonstrated to show poor performance over additive white Gaussian noise (AWGN) channel \cite{cammerer2017sparse}. Pruning methods for polar factor graph are consequently proposed in \cite{cammerer2017sparse} to perform efficient polar decoding with LDPC-like manner. The sparse graph after using pruning techniques is shown as Fig. \ref{fig:sparse_tanner}. For more details, we refer the readers to \cite{goela2010lp, cammerer2017sparse}.
	
%
%

	\section{Proposed Sparse Neural Network Decoder} \label{sec:proposed_snnd}
	\subsection{Trellis Construction of Sparse Neural Network Decoder}
	

	The trellis of proposed SNND is constructed based on the sparse polar Tanner graph in \cite{cammerer2017sparse}. More specifically, proposed SNND is a deep feed-forward neural network similar to the structure of \cite{nachmani2016learning}. The nodes of each hidden layer represent corresponding edges in the Tanner graph.
	
	\begin{figure*}[t]
		\centering
		\includegraphics[width=0.75\linewidth]{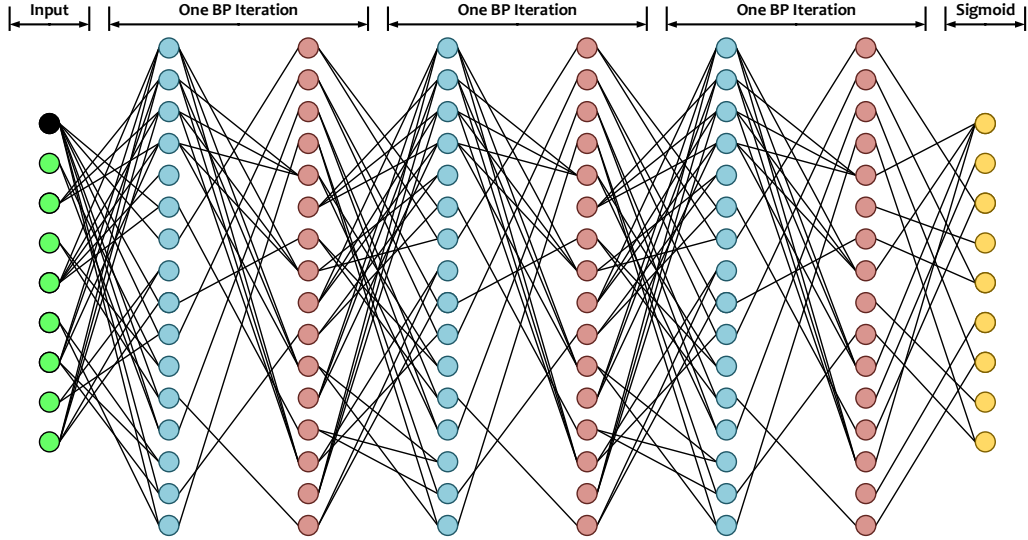}
		\label{fig:sparse_nnd}
		\caption{Sparse neural network decoder (SNND) for (8,4) polar code with 6 hidden layers.}
		\label{fig:nnd}
	\end{figure*}

	\begin{equation}\label{eq:sparse_polar_8_4}
		\mathbf{H}=
		\begin{bmatrix}
		0 & 0 & 1 & 0 & 1 & 0 & 1 & 0 & 1\\
		1 & 0 & 0 & 1 & 1 & 0 & 0 & 0 & 0\\
		1 & 0 & 0 & 0 & 0 & 0 & 0 & 1 & 1\\
		1 & 1 & 1 & 0 & 0 & 0 & 0 & 0 & 0\\
		1 & 0 & 0 & 0 & 0 & 1 & 1 & 0 & 0
		\end{bmatrix}
	\end{equation}
	
	The trellis construction of (8,4) polar SNND is given as an example. The conventional factor graph associated with generator matrix $\mathbf{G}_{8} = \mathbf{F}^{\otimes 3}$ is first converted into the LDPC-like Tanner graph (see Fig. \ref{fig:dense_tanner}) consisting of VNs and CNs \cite{goela2010lp}. Then we use the pruning techniques of \cite{cammerer2017sparse} to reduce the number of edges, converting the dense graph into a sparse Tanner graph (see Fig. \ref{fig:sparse_tanner}). The resulting parity-check matrix $\mathbf{H}$ is shown in Eq. (\ref{eq:sparse_polar_8_4}). Note that the sparse Tanner graph is slightly different from LDPC codes since a portion of edges from VNs to CNs are not removed \cite{cammerer2017sparse} (\textit{black VN} in Fig. \ref{fig:sparse_tanner}).
	
	Next, the bipartite sparse Tanner graph is unfolded and converted into the feed-forward neural network in Fig. \ref{fig:nnd}. Assume that we have an ($N, K$) polar code on sparse Tanner graph with total $E$ edges, $N_{v}$ VNs, and $T$ iterations in the sparse Tanner graph. The associated SNND has $2T$ hidden layers. For the input layer with $N_{v}$ nodes, the initial LLRs of received channel output are fed into the last $N$ nodes. The number of nodes in each hidden layer equals to the edges $E$ and each hidden node denotes the soft message propagated over corresponding edge. The final $N_{v}$ outputs are activated by the sigmoid function.

	\subsection{Decoding Process}\label{subsec:decoding}
	
	Let $\mathbf{x}=(x_{1},...,x_{N})$ be the transmitted codeword with systematic encoding \cite{arikan2011systematic} and $\mathbf{y}=(y_{1},...,y_{N})$ be the received channel output. The input size of SNND is slightly larger than $N$ since part of VNs are not removed. The initial LLR of the $v$-th node in input layer is computed as the following equation:
	\begin{equation}\label{eq:llr}
		L_{v} =\begin{cases}
		\qquad\qquad 0, & 1 \le v \le N_{v}-N, \\
		\log{\dfrac{P(x_{j}=0|y_{j})}{P(x_{j}=1|y_{j})}}, & N_{v}-N+1 \le v \le N_{v},
		\end{cases}
	\end{equation}
	where we have $j = v - (N_{v} - N)$.
	
	The standard SPA can be used to construct polar codes over Tanner graphs as \cite{nachmani2016learning,nachmani2018near}. But the computational complexity of SPA is prohibitive due to the hyperbolic trigonometric function and multiplication. \cite{nachmani2018deep} demonstrates that NND constructed by MS decoding can also achieve promising performance compared with SPA. Therefore we use the simplified MS decoding to define the two types of basic neurons in SNND see Fig. \ref{fig:nnd}. Each neuron represents the associated edge in Tanner graph. The odd layer $i$ only contains neurons without any parameters. The updating function is the MS approximation:
	\begin{equation}\label{eq:snnd_odd}
		x_{i,e=(c,v)} = \prod_{e'=(v', c), v'\neq v} \text{sign} (x_{i-1, e'}) \cdot \min(| x_{i-1, e'} |),
	\end{equation}
	where $e'=(v', c)$ denotes the set of VNs $v'$ connected to CNs $c$.	
	
	The even hidden layer $i$ only contains neurons that assign weights to incoming messages as follows:
	\begin{equation}\label{eq:snnd_even}
		x_{i, e=(v,c)} = L_{v} + \sum_{e'=(c', v), c'\neq c} w_{i,e,e'} x_{i-1,e'}.
	\end{equation}
	
	The output layer squashes the final weighted soft messages to the range $[0,1]$ as follows:		
	\begin{equation}\label{eq:snnd_out}
		o_{v} = \sigma (L_{v} + \sum_{e'=(c', v)} w_{2L+1,v,e'}x_{2L,e'}),
	\end{equation}
	where $\sigma(x)=(1+e^{-x})^{-1}$ is the sigmoid function. Note that the sigmoid function is only applied to the output layer during training phase. For simplicity, the feed-forward SNND is defined as SNND-FF.

	\subsection{Optimizing with Single Weight}
	The decoding complexity of SNND is significantly reduced compared to the original SPA. However, the required number of weights is still large. For the RNNs in \cite{nachmani2018deep}, the weights of edges are shared within each BP iteration. Besides, the RNN structure is easier to optimize compared to the feed-forward counterparts. There is still some redundancy for the RNN structure. We further reduce the required number of weights to just one as follows:
	\begin{equation}\label{eq:snnd_odd_single}
		x_{i, e=(v,c)} = L_{v} + \sum_{e'=(c', v), c'\neq c} w'  x_{i-1,e'},
	\end{equation}
	where $w'$ denotes the unified weight for all edges from CNs to VNs. $w'$ is also applied to the final output in Eq. (\ref{eq:snnd_out}).
	
	The optimization is easier and the optimal parameter $w^{*}$ is given by $w$ that results in the minimum loss:
	\begin{equation}\label{eq:opt_target}
		w^{*} = \argmin_{w'} \mathcal{L}(\bm{x}, \bm{o}).
	\end{equation}
	
	\subsection{Training of Sparse Neural Network Decoder}
	The cross entropy function is adopted to express the evaluate the loss between neural network output $\mathbf{o}$ and the transmitted codeword $\mathbf{x}$:
	\begin{equation}\label{eqn_loss}
		\mathcal{L}(\mathbf{x}, \mathbf{o}) = -\dfrac{1}{N} \sum _{i=N'} ^{N_{v}} x_{i}\log(o_{i}) + (1-x_{i})\log(1-o_{i}),
	\end{equation}
	where $o_{i}$, $x_{i}$ denote the $i$-th bit of SNND outputs and the $i$-th bit of transmitted codeword, respectively. The last $N$ bits are calculated and $N' = N_{v}-N+1$.
	
	The parameter space of SNND is determined by the total edges in corresponding sparse Tanner graph and the iteration number. Hence, the optimization space grows larger when the code length and the iteration increase. A good parameter initialization can boost the convergence of training. \cite{Goodfellow-et-al-2016} suggests to initialize the parameters with a normal distribution. But the standard normal distribution is unable to guarantee a quick convergence. We initialize the parameters of feed-forward SNND to a normal distribution with mean $\mu=1$ and a small variance $\sigma$ in the experiment while the SNND with single weight is initialized to one.

	\section{Experiment} \label{sec:experiment}
	\subsection{Setup}
	The SNND is implemented on deep learning library \textit{PyTorch}. We use mini-batch stochastic gradient descent (SGD) with Adam \cite{kingma2014adam} algorithm to optimize the neural network. The learning rate Lr is set to $0.001$. AWGN channel and binary phase-shift keying (BPSK) modulation with SNR range $1$ to $4$ are considered. As in \cite{nachmani2016learning}, the training set consists of all zero codeword and the mini-batch size is $120$ ($30$ samples per SNR). The parameters are initialized with normal distribution $\mathcal{N}\sim (\mu=1, \sigma=0.1)$. Zero-value messages in the SNND will make the CN-to-VN messages in Eq. (\ref{eq:snnd_odd}) to be zero, which hinders the message propagation. To avoid this issue, the result of sign operation for a zero value is defined as $1$.
	
	

	\subsection{Results}
\begin{figure}[ht]
		\centering
		\includegraphics[width=.9\linewidth]{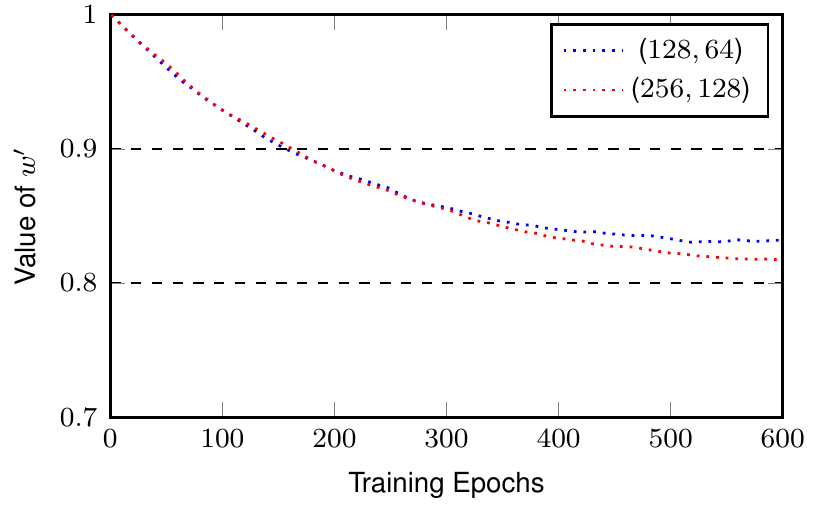}
		\caption{Evolution of trained weight $w'$ of SNND.}
		\label{fig:training_loss}
	\end{figure}

	We train two types of SNND: SNND-FF and SNND with single weight. Both of them are unfolded to $10$ iterations, corresponding to $20$-layered neural networks. Each network is trained for $600$ epochs. Fig. \ref{fig:training_loss} illustrates the trend of trained unified weight $w'$ on ($128,64$) and ($256,128$) polar codes. The trained optimal $w^{*}$ for ($128,64$) code finally converges to $0.83$ while the value of $w^{*}$ for ($256,128$) code is closed to $0.82$.
	
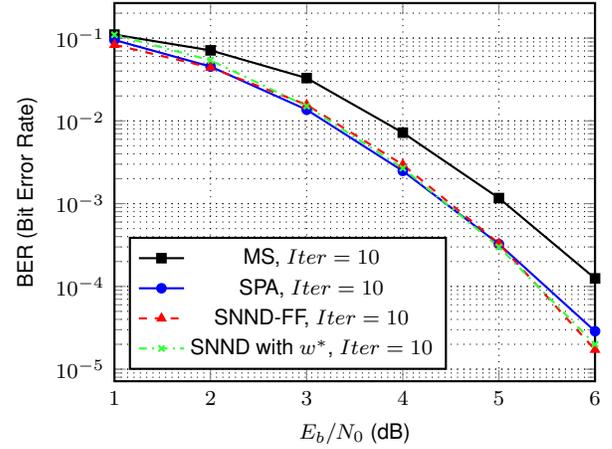
\begin{figure}[ht]
		\centering
		\begin{tikzpicture}[font=\sffamily\footnotesize]
		\begin{semilogyaxis}[
		height=0.28\textheight,
		width=0.9\linewidth,
		xmin = 1, xmax = 6,
		grid=both,
		xlabel=$E_b / N_0$ (dB),
		ylabel=BER (Bit Error Rate),
		legend pos = south west,
		line width=0.8pt,
		]
		
		\addplot[color=black, mark=square*, mark options={scale=0.8, solid}] coordinates {
			(1, 0.11035156 )
			(2, 0.07109375 )
			(3, 0.03291016 )
			(4, 0.00720486 )
			(5, 0.00116801 )
			(6, 0.00012500 )
		};
		\addlegendentry{MS, $Iter = 10$}

		\addplot[color=blue, mark=*, mark options={scale=0.8, solid}] coordinates {
			(1, 0.09498875000 )
			(2, 0.04529968750 )
			(3, 0.01364234375 )
			(4, 0.00248468750 )
			(5, 0.00032453125 )
			(6, 0.00002875000 )
		};
		\addlegendentry{SPA, $Iter = 10$}

		\addplot[color=red, dashed, mark=triangle*, mark options={scale=0.8, solid}] coordinates {
			(1, 0.08300781 )
			(2, 0.04414063 )
			(3, 0.01562500 )
			(4, 0.00295351 )
			(5, 0.00033044 )
			(6, 0.00001719 )
		};
		\addlegendentry{SNND-FF, $Iter = 10$}
		
		\addplot[color=green!80, dash dot, mark=x, mark options={scale=0.8, solid}] coordinates {
			(1,   0.109876 )
			(2,   0.053931 )
			(3,   0.015045 )
			(4,   0.002687 )
			(5,   0.000300 )
			(6,   0.000020 )
		};
		\addlegendentry{SNND with $w^{*}$, $Iter = 10$}
		
		\end{semilogyaxis}
		\end{tikzpicture}
		\caption{BER comparison of trained SNNDs and different decoding schemes on ($128,64$) polar code with $10$ iterations.}
		\label{fig:SNND_128_64_ber_iter_10}
	\end{figure}

\begin{figure}[ht]
		\centering
		\begin{tikzpicture}[font=\sffamily\footnotesize]
		\begin{semilogyaxis}[
		height=0.28\textheight,
		width=0.9\linewidth,
		xmin=1, xmax=6,
		grid=both,
		xlabel=$E_b / N_0$ (dB),
		ylabel=BER (Bit Error Rate),
		legend pos = south west,
		line width=0.8pt,
		]		
		
		\addplot[color=black, mark=square*, mark options={scale=0.8, solid}] coordinates {
			(1,   0.137160 )
			(2,   0.072059 )
			(3,   0.026597 )
			(4,   0.007336 )
			(5,   0.001455 )
			(6,   0.000200 )
		};
		\addlegendentry{MS, $Iter = 10$}

		\addplot[color=blue, mark=*, mark options={scale=0.8, solid}] coordinates {
			(1,   0.103847 )
			(2,   0.049834 )
			(3,   0.017938 )
			(4,   0.004360 )
			(5,   0.000708 )
			(6,   0.000096 )
		};
		\addlegendentry{SPA, $Iter = 10$}

		\addplot[color=red, dashed, mark=triangle*, mark options={scale=0.8, solid}] coordinates {
			(1,   0.091837 )
			(2,   0.053382 )
			(3,   0.019126 )
			(4,   0.005013 )
			(5,   0.000886 )
			(6,   0.000115 )
		};
		\addlegendentry{SNND-FF, $Iter = 10$}
		
		\addplot[color=green!80, dash dot, mark=x, mark options={scale=0.8, solid}] coordinates {
			(1,   0.114523 )
			(2,   0.052227 )
			(3,   0.018959 )
			(4,   0.005219 )
			(5,   0.000952 )
			(6,   0.000116 )
		};
		\addlegendentry{SNND with $w^{*}$, $Iter = 10$}
		
		\end{semilogyaxis}
		\end{tikzpicture}
		\caption{BER comparison of trained SNNDs and different decoding schemes on ($256,128$) polar code with $10$ iterations.}
		\label{fig:SNND_256_128_ber_iter_10}
	\end{figure}
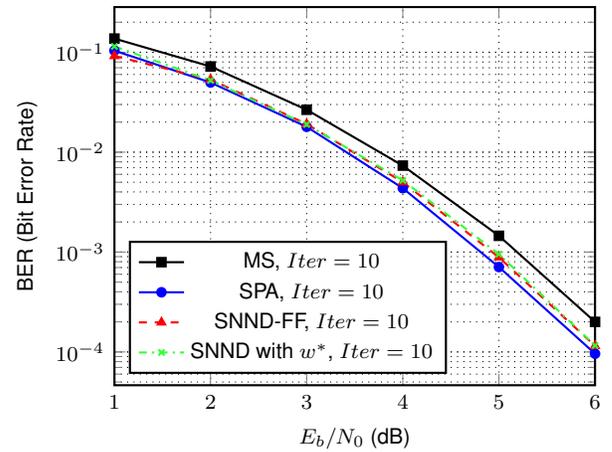

	The BER performance is evaluated with four types of decoding schemes: 1) SPA on sparse Tanner graph \cite{cammerer2017sparse}, 2) MS algorithm on sparse Tanner graph, 3) Proposed SNND-FF, 4) Proposed SNND with single weight. Fig. \ref{fig:SNND_128_64_ber_iter_10} illustrates the BER results for two trained SNNDs on ($128, 64$) polar code. The SNND-FF equivalent to $10$ iterations achieves almost the same performance with SPA and has an improvement of about $0.1$ to $0.5$ dB over MS decoding. The gap between SNND-FF and SNND with single weight is negligible. Fig. \ref{fig:SNND_256_128_ber_iter_10} shows the BER comparison on ($256,128$) polar code. The SNNDs have about $0.15$ dB gain over MS decoding and have less than $0.1$ dB performance degradation in high SNR region compared with SPA.
	
	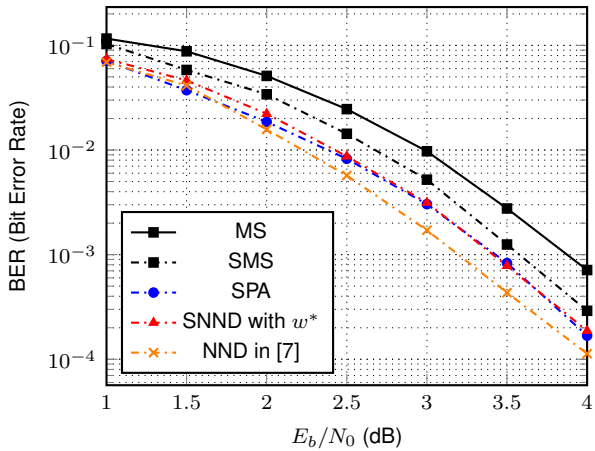
\begin{figure}[ht]
	\centering
	\begin{tikzpicture}[font=\sffamily\footnotesize]
		\begin{semilogyaxis}[
		height=0.28\textheight,
		width=0.9\linewidth,
		xmin = 1, xmax = 4,
		grid=both,
		xlabel=$E_b / N_0$ (dB),
		ylabel=BER (Bit Error Rate),
		legend pos = south west,
		line width=0.8pt,
		]
	
		\addplot[color=black, mark=square*, mark options={scale=0.8, solid}] coordinates {
			(1,   0.116154296 )
			(1.5, 0.087798438 )
			(2,   0.051101562 )
			(2.5, 0.024591438 )
			(3,   0.009728515 )
			(3.5, 0.002762038 )
			(4,   0.000712382 )
		};
		\addlegendentry{MS}

		\addplot[color=black, dash dot, mark=square*, mark options={scale=0.8, solid}] coordinates {
			(1,   0.1030398438 )
			(1.5, 0.0582550000 )
			(2,   0.0340656250 )
			(2.5, 0.0142940008 )
			(3,   0.0052085938 )
			(3.5, 0.0012498438 )
			(4,   0.0002906250 )
		};
		\addlegendentry{SMS}

		\addplot[color=blue, mark=*, dash dot, mark options={scale=0.8, solid}] coordinates {
			(1,   0.071019 )
			(1.5, 0.037269 )
			(2,   0.018570 )
			(2.5, 0.008238 )
			(3,   0.003035 )
			(3.5, 0.000831 )
			(4,   0.000168 )
		};
		\addlegendentry{SPA}

		\addplot[color=red, dash dot, mark=triangle*, mark options={scale=0.8, solid}] coordinates {
			(1,   0.074001 )
			(1.5, 0.046255 )
			(2,   0.022201 )
			(2.5, 0.008696 )
			(3,   0.003140 )
			(3.5, 0.000782 )
			(4,   0.000187 )
		};
		\addlegendentry{SNND with $w^{*}$}

		\addplot[color=orange, dash dot, mark=x, mark options={scale=1.2, solid}] coordinates {
			(1,   0.069048 )
			(1.5, 0.041264 )
			(2,   0.015726 )
			(2.5, 0.005700 )
			(3,   0.001711 )
			(3.5, 0.000433 )
			(4,   0.000113 )
		};
		\addlegendentry{NND in \cite{xu2017improved}}
		
		\end{semilogyaxis}
		\end{tikzpicture}
		\caption{BER comparison for various decoding schemes on ($128,64$) polar code with $50$ iterations.}
		\label{fig:ber_128_64_ber}
	\end{figure}

	\begin{figure}[ht]
		\centering
		\begin{tikzpicture}[font=\sffamily\footnotesize]
		\begin{semilogyaxis}[
		height=0.28\textheight,
		width=0.9\linewidth,
		xmin = 1, xmax = 4,
		grid=both,
		xlabel=$E_b / N_0$ (dB),
		ylabel=BER (Bit Error Rate),
		legend pos = south west,
		line width=0.8pt,
		]
		
		\addplot[color=black, mark=square*, mark options={scale=0.8, solid}] coordinates {
			(1,   0.14640000 )
			(1.5, 0.08887300 )
			(2,   0.04983281 )
			(2.5, 0.016081 )
			(3,   0.00427000 )
			(3.5, 0.000801 )
			(4,   0.000139843 )
		};
		\addlegendentry{MS}

		\addplot[color=black, dash dot, mark=square*, mark options={scale=0.8, solid}] coordinates {
			(1,   0.116032 )
			(1.5, 0.063254 )
			(2,   0.027342 )
			(2.5, 0.007355 )
			(3,   0.001551 )
			(3.5, 0.000263 )
			(4,   0.000043 )
		};
		\addlegendentry{SMS}

		\addplot[color=blue, mark=*, dash dot, mark options={scale=0.8, solid}] coordinates {
			(1,   0.066142 )
			(1.5, 0.032655 )
			(2,   0.013354 )
			(2.5, 0.004114 )
			(3,   0.000896 )
			(3.5, 0.000152 )
			(4,   0.0000271094 )
		};
		\addlegendentry{SPA}

		\addplot[color=red, dash dot, mark=triangle*, mark options={scale=0.8, solid}] coordinates {
			(1,   0.073794 )
			(1.5, 0.039051 )
			(2,   0.013988 )
			(2.5, 0.004001 )
			(3,   0.000765 )
			(3.5, 0.000132 )
			(4,   0.0000198 )
		};
		\addlegendentry{SNND with $w^{*}$}

		\addplot[color=orange, dash dot, mark=x, mark options={scale=1.2, solid}] coordinates {
			(1,   0.061048 )
			(1.5, 0.026497 )
			(2,   0.0091383 )
			(2.5, 0.002227 )
			(3,   0.000463 )
			(3.5, 0.000077 )
			(4,   0.000012 )
		};
		\addlegendentry{NND in \cite{xu2017improved}}
		
		\end{semilogyaxis}
		\end{tikzpicture}
		\caption{BER comparison for various decoding schemes on ($256,128$) polar code with $50$ iterations.}
		\label{fig:ber_256_128_ber}
	\end{figure}
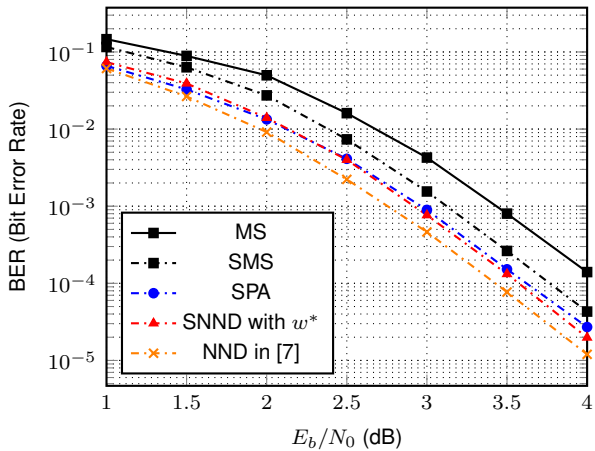
	
	We also compare the SNND with other decoding algorithms. The scaled min-sum (SMS) and neural network decoder (NND) in \cite{xu2017improved} are considered. The scaling factor of SMS equals to $0.9375$, which is suggested in \cite{yuan2014early}. The NND is trained by unfolding to $10$ iterations and tested with $50$ iterations. After increasing the iteration number, the SNND with single weight $w^{*}$ can obtain better performance. Fig. \ref{fig:ber_128_64_ber} shows the performance comparison for various decoding schemes with $50$ iterations. The SNND with trained $w^{*}$ achieves comparative performance with SPA and outperforms SMS and MS by about $0.1$ dB and $0.4$ dB, respectively. Due to pruning of some connections, the SNND has $0.1$ dB degradation compared with polar NND in \cite{xu2017improved}. The similar results can also be observed on ($256,128$) polar code in Fig. \ref{fig:ber_256_128_ber}.

	\subsection{Complexity and Latency Analysis}\label{sec:complexity_analysis}
	The latency and complexity of proposed SNND are both reduced compared to NND in \cite{xu2017improved} and the original polar BP decoding with SMS \cite{yuan2014early}. The original polar BP \cite{arikan2008performance} will consecutively activate $2\log_{2}N$ stages during left-to-right and right-to-left propagations, resulting a latency of $2\log_{2}N$ time steps for each iteration. Besides, $N$ multiplications, $N$ additions, and $N$ comparisons are required for each stage. Hence, the total complexity of one iteration is $\mathcal{O}(2N\log_{2}N)$.
	
	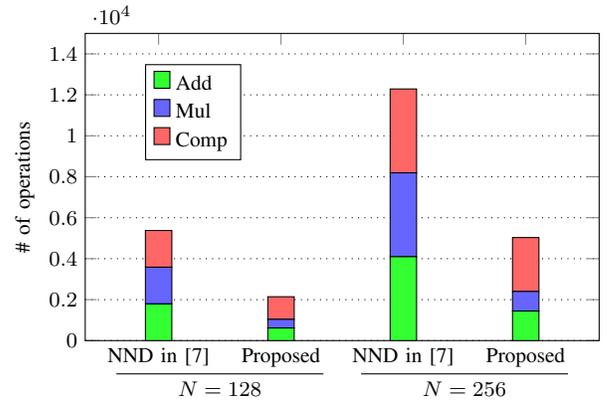
\begin{figure}[ht]
		\centering
		\begin{tikzpicture}
		\footnotesize
		\begin{axis}[
		width=0.95\linewidth,
		height = 0.24\textheight,
		ylabel={\# of operations},
		ymajorgrids = true,
		xtick=data,
		xticklabels={NND in \cite{xu2017improved}, Proposed, NND in \cite{xu2017improved}, Proposed},
		ytick distance={0.2*1e4},
		enlarge y limits=false,
		ymin=0, ymax=1.5e4,
		enlarge x limits=0.2,
		ybar stacked,
		bar width=10pt,
		legend style={
			cells={anchor=west},
			legend columns=1,
			at={(0.21,0.9)},
			anchor=north,
			/tikz/every even column/.append style={column sep=0.2cm}
		},
		draw group line={[index]4}{1}{$N = 128$}{-3.5ex}{16pt},
		draw group line={[index]4}{2}{$N = 256$}{-3.5ex}{16pt}
		]
		
		\addplot [fill=green!80]	table[x index=0,y index=1] \datatable;
		\addplot [fill=blue!60]		table[x index=0,y index=2] \datatable;
		\addplot [fill=red!60,
		]		
		table[x index=0, y index=3] \datatable;
		\legend{Add, Mul, Comp}
		\end{axis}
		\end{tikzpicture}
		\caption{Complexity comparison of one iteration for proposed SNND with single weight and polar NND in \cite{xu2017improved}.}
		\label{fig:complexity_comparison}
	\end{figure}

	The complexity of SNND with single weight is determined by corresponding $\mathbf{H}$ matrix. Each CN with $d_{c}$ incoming messages requires $2d_{c}$ comparisons to find the minimum and $2$nd minimum value. $2$ multiplications are needed to compute the outgoing messages. Each VN with $d_{v}$ incoming messages requires $d_{v}$ additions. Note that the sign operation of CNs is omitted since its complexity is very low. SNND implements a LDPC-like flooding pattern with high parallelism. The latency for each iteration equals to $2$, which is independent with code length. Hence, the latency reduction is $\log_{2}N$ compared with the NND \cite{xu2017improved} and original BP. Fig. \ref{fig:complexity_comparison} gives the number of three types of operations (addition, multiplication and comparison) for proposed SNND with single weight and NND in \cite{xu2017improved}. The SNND can reduce about $60 \%$ operations on the two mentioned code lengths.

	\section{Conclusion}\label{sec:conclusion}
	In this work, we propose a fully parallel neural network decoder for polar codes. The SNND is constructed from the sparse Tanner graph of polar codes \cite{cammerer2017sparse}. Then the weights of SNND are dramatically reduced to just one by using single parameter. Deep learning techniques are utilized to optimize the networks. Compared with conventional BP, the results show that SNND achieves competitive BER performance. Moreover, the complexity and latency are much lower according to the analysis. Our future work will focus on further improvements of SNND using other decoding methods, such as \cite{nachmani2018deep} or \cite{nachmani2018near}.
	
	\section*{Acknowledgement}
	This work is supported in part by NSFC under grants $61871115$ and $61501116$, Jiangsu Provincial NSF for Excellent Young Scholars under grant BK$20180059$, Huawei HIRP Flagship under grant YB$201504$, the Fundamental Research Funds for the Central Universities, the SRTP of Southeast University, State Key Laboratory of ASIC \& System under grant $2016$KF$007$, ICRI for MNC, and the Project Sponsored by the SRF for the Returned Overseas Chinese Scholars of MoE.
	
	\footnotesize
	\bibliographystyle{IEEEtran}
	\bibliography{Bib}
	
\end{document}